\documentclass[
reprint,           
onecolumn,
superscriptaddress,
amsmath,           
amssymb,           
aps,               
prd,               
notitlepage,       
longbibliography,  
floatfix,          
nofootinbib,
]{revtex4-1}
\usepackage{amsmath}
\allowdisplaybreaks[4]
\usepackage{cancel}
\usepackage{extarrows}
\usepackage{tensor}     
\usepackage[caption = false]{subfig} 
\usepackage[
colorlinks=true,      
citecolor=blue,         
linkcolor=blue,         
urlcolor=blue           
]{hyperref}             
\usepackage{bm}         %
\usepackage{xcolor}     %
\usepackage{lipsum}
\usepackage{color}      
\usepackage[utf8]{inputenc} 
\usepackage[section]{placeins} 
\usepackage{appendix}
\usepackage{units}
\usepackage{orcidlink}
\usepackage{tabularx}
\usepackage{adjustbox}
\usepackage{graphicx}
\newcommand{\nc}{\newcommand*} 

\usepackage{float}

\usepackage{placeins}
\nc{\tr}{\rm{tr}}
\nc{\sig}{\sigma}
\nc{\om}{\omega}
\nc{\bt}{\beta}
\nc{\nb}{\nabla}
\nc{\eps}{\epsilon}
\nc{\epsS}{\epsilon_{\rm{S}}}
\nc{\epsV}{\epsilon_{\rm{V}}}
\nc{\epsT}{\epsilon_{\rm{T}}}
\nc{\osc}{\rm{osc}}
\nc{\DM}{\rm{TF}}
\nc{\FP}{\rm{FP}}
\nc{\eV}{\rm{eV}}
\nc{\figurewidth}{3.2in}
\nc{\xbar}{\bar{x}}
\nc{\rhoeq}{\rho_{\mathrm{eq}}}
\nc{\zeq}{z_{\mathrm{eq}}}
\nc{\tla}{\tilde{\lambda}}
\nc{\dt}{\delta}
\nc{\Dt}{\Delta}
\nc{\vj}{\hat{j}}
\nc{\vl}{\hat{l}}
\nc{\hx}{\hat{x}}
\nc{\hy}{\hat{y}}
\nc{\bj}{\bm{j}}
\nc{\mJ}{\mathcal{J}}
\nc{\mP}{\mathcal{P}}
\nc{\lbd}{\lambda}
\nc{\Msun}{M_\odot}
\nc{\app}{\approx}
\nc{\av}[1]{\langle #1 \rangle}
\nc{\eq}[1]{Eq.~\eqref{#1}}
\nc{\al}{\alpha}
\nc{\Xstar}{X_{\ast}}
\nc{\seq}{\sigma_{\mathrm{eq}}}
\nc{\fpbh}{f_{\mathrm{pbh}}}
\nc{\vth}{\hat{\theta}}
\nc{\vla}{\hat{\lambda}}
\nc{\vd}{\hat{d}}
\nc{\Mmin}{M_{\mathrm{min}}}
\nc{\rmd}{\mathrm{d}}
\nc{\mmin}{{m_{\mathrm{min}}}}
\nc{\mmax}{{m_{\mathrm{max}}}}
\nc{\mR}{\mathcal{R}}
\nc{\tmR}{\tilde{\mathcal{R}}}
\nc{\s}{\sigma}
\nc{\ogw}{\Omega_{\mathrm{GW}}}
\nc{\addref}{[\textcolor{red}{add ref}] }
\nc{\Om}{\Omega}
\nc{\gm}{\gamma}
\nc{\Gm}{\Gamma}
\nc{\kp}{\kappa}
\nc{\hbe}{\hat{\mathbf{e}}}
\nc{\gpcyr}{\mathrm{Gpc}^{-3}\,\mathrm{yr}^{-1}}
\nc{\Eq}[1]{Eq.~\eqref{#1}}
\nc{\Fig}[1]{Fig.~\ref{#1}}
\nc{\Table}[1]{Table~\ref{#1}}
\nc{\lvc}{LIGO/Virgo} 
\nc{\Sec}[1]{Sec.~\ref{#1}}
\nc{\eg}{\textit{e.g.~}}
\nc{\SNR}{\mathrm{SNR}}
\nc{\be}{\mathbf{\epsilon}}
\nc{\bn}{\mathbf{n}}
\nc{\bx}{\mathbf{x}}
\nc{\bk}{\mathbf{k}}
\nc{\bd}{\mathbf{d}}
\nc{\ba}{\mathbf{a}}
\nc{\bp}{\mathbf{p}}
\nc{\bnu}{\mathbf{\nu}}
\nc{\uni}{\mathrm{U}}
\nc{\logu}{\operatorname{\mathrm{log-U}}}
\nc{\RN}{\mathrm{RN}}
\nc{\BN}{\mathrm{BN}}
\nc{\GN}{\mathrm{GN}}
\nc{\mcN}{\mathcal{N}}
\nc{\GWB}{\mathrm{GW}}
\nc{\yr}{\mathrm{yr}}
\nc{\Am}{\mathcal{A}}
\nc{\Dm}{\mathcal{D}}
\nc{\Hm}{\mathcal{H}}
\nc{\sovast}{Soviet Ast.}
\nc{\hosc}{h_{\mathrm{osc}}}
\nc{\Posc}{\Psi_{\mathrm{osc}}}

\nc{\mrm}{\mathrm}
\nc{\BE}{B\scriptsize{AYES}\normalsize{E}\scriptsize{PHEM}\normalsize  }

\def\({\left(}
\def\){\right)}
\def\[{\left[}
\def\]{\right]}

\def\e{\begin{equation}}
\def\q{\end{equation}}
\def\m{\begin{eqnarray}}
\def\n{\end{eqnarray}}
\nc{\red}[1]{\textcolor{red}{#1}}
\begin{document}

\title{Pulsar Timing Response and Spatial Correlations of Shear Modes
in Torsionless Palatini Spacetime}     

\author{Tian-Shi Li}
\affiliation{Center for Gravitation and Cosmology, College of Physical Science and Technology, Yangzhou University, Yangzhou, 225009, China}

\author{Yu-Mei Wu\orcidlink{0000-0002-9247-5155}}
\email{ wuyumei@yzu.edu.cn} 
\affiliation{Center for Gravitation and Cosmology, College of Physical Science and Technology, Yangzhou University, Yangzhou, 225009, China}

\author{Chang Liu 
}
\email[]{liuchang@yzu.edu.cn}
\affiliation{Center for Gravitation and Cosmology, College of Physical Science and Technology, Yangzhou University, Yangzhou, 225009, China}


\begin{abstract}
In the nanohertz band, spatial correlations between pulsar timing
residuals provide a key observable for characterizing stochastic
gravitational-wave backgrounds and probing their polarization content.
In torsionless Palatini spacetime, nonmetricity can generate two
additional shear modes, referred to as the shear-$x$ and shear-$y$
modes. In this work, we investigate the pulsar timing response produced
by the shear-induced motions of the emitting pulsar and the receiving
Earth. Assuming that the electromagnetic field is minimally coupled to
the physical metric and that the Earth and pulsar possess non-negligible
effective hypermomentum responses, we derive the single-pulsar redshift,
the frequency-domain two-point correlation function, and the
corresponding spatial correlation. For a stationary and isotropic
stochastic shear background, the normalized overlap reduction function
(ORF) for distinct pulsars reduces in the short-wavelength limit to the
pure dipolar form $\Gamma_{ab}^{\mathrm{sh}}(\zeta)=\cos\zeta$, where
$\zeta$ is the angular separation between the two pulsars. The same
dipolar correlation can also be produced by isotropic Solar System
ephemeris errors, leading to a spatial degeneracy between the two
signals. This degeneracy highlights the importance of information
beyond the angular correlation for identifying Palatini shear
signatures in PTA data.
\end{abstract}

	
\maketitle

\section{Introduction}
Pulsar timing arrays (PTAs) search for gravitational-wave-induced timing residuals in the nanohertz band by monitoring the pulse times of arrival from millisecond pulsars over many years. In addition to the low-frequency power measured for individual pulsars, correlations between the timing residuals of different pulsars encode the characteristic spatial pattern of a stochastic gravitational-wave background (SGWB) and therefore provide a key means of distinguishing it from generic noise processes \cite{Sazhin:1978myk,Detweiler:1979wn,foster1990constructing}. For a stationary, Gaussian, isotropic, and unpolarized tensor
background in general relativity, this correlation depends only on the angular separation between pulsars and is described by the well-known Hellings--Downs (HD) curve \cite{hellings1983upper}. 

After several decades of timing observations, PTAs have made substantial progress in the search for a nanohertz SGWB. 
The North American Nanohertz Observatory for Gravitational Waves (NANOGrav) first reported a common-spectrum red-noise process across pulsars in its 12.5-year data set \cite{arzoumanian2020nanograv}. Similar common signals were subsequently reported by the Parkes Pulsar Timing Array (PPTA)  \cite{goncharov2021evidence}, the European Pulsar Timing Array (EPTA)  \cite{chen2021common}, and the International Pulsar Timing Array (IPTA) in its second data release \cite{antoniadis2022international}. In 2023, NANOGrav, the EPTA, together with the Indian Pulsar Timing Array (InPTA), the PPTA, and the Chinese Pulsar Timing Array (CPTA), reported evidence for the characteristic HD spatial correlation in their latest data releases \cite{agazie2023nanograv,antoniadis2023second,reardon2023search,xu2023searching}. These results provide compelling observational evidence for the existence of a nanohertz SGWB.

Despite its remarkable success in describing gravitational phenomena and passing a wide range of experimental tests, GR remains difficult to reconcile with quantum theory and does not by itself provide an explanation for dark matter and dark energy. These open questions have motivated numerous extensions and alternatives to GR. In theories beyond GR, gravitational waves (GWs) may possess additional polarization modes or propagation properties different from those predicted by GR, leading to spatial correlations distinct from the standard HD curve. PTA observations have therefore been used to search for scalar and vector stochastic backgrounds and to constrain additional GW polarizations \cite{lee2008pulsar,
Chamberlin:2011ev,
Cornish:2017oic,
Chen:2021wdo,
NANOGrav:2021ini,
wu2022constraining,
NANOGrav:2023ygs,
chen2024search}, while modified GW propagation and its effects on PTA spatial correlations have also been investigated \cite{Lee:2010cg,
Bernardo:2023mxc,
Wang:2023div,
Wu:2023rib,
Bernardo:2023zna,
Liang:2024mex}. The spatial correlation between pulsar pairs therefore provides a useful observable for testing the GW predictions
of GR and alternative theories \cite{Gair:2015hra,NANOGrav:2024tnd}.

Most studies of alternative GW polarizations in PTAs have focused on theories in which the gravitational response is described within a metric framework. A broader geometric setting is provided by metric--affine gravity, in which the metric and affine connection are treated as independent variables and the connection can contain torsion and nonmetricity in addition to curvature \cite{hehl1995metric,vitagliano2011dynamics,Iosifidis:2019dua,
Rigouzzo:2022yan,heisenberg2024review}. In such theories, the independent connection can affect the motion of matter when the matter sector has a non-negligible hypermomentum response \cite{Neeman:1996zcr,Puetzfeld:2007ye,Puetzfeld:2007hr,
puetzfeld2014equations,iosifidis2024motion}. GW polarizations have also been investigated in
specific Palatini theories, including generalized Brans--Dicke and
Palatini--Horndeski gravity
\cite{Lu:2020eux,Dong:2021jtd}. In torsionless Palatini spacetime, recent gauge-invariant analyses have shown that nonmetricity can generate two additional shear components, referred to as the shear-$x$ and shear-$y$ modes, whose particle-motion patterns differ from those of the conventional Riemannian polarizations \cite{dong2025new}. While their geometric origin and particle-motion properties have been investigated, their signatures in pulsar timing have not yet been explored.

In this work, we investigate the pulsar timing response produced by these Palatini shear modes. We assume that the electromagnetic
field is minimally coupled to the metric and has no direct coupling to the independent affine connection, while allowing the matter
degrees of freedom forming the pulsar and Earth to possess a non-negligible effective hypermomentum response. Under
these assumptions, the shear sector produces an additional timing signal through the local motions of the emitting pulsar and the
receiving Earth, rather than through a direct coupling of the independent connection to the electromagnetic field.

We derive the single-pulsar redshift response, the frequency-domain
two-point correlation function for pulsar pairs, and the corresponding
spatial correlation for an isotropic stochastic shear background.
Denoting the angular separation between two pulsars by $\zeta$, the
normalized overlap reduction function (ORF) for distinct pulsars
reduces in the short-wavelength limit to the dipolar form
$\Gamma^{\mathrm{sh}}(\zeta)=\cos\zeta$.
Section~\ref{sec:shear} reviews the geometric origin and
particle-motion properties of the shear modes in torsionless Palatini
spacetime. Section~\ref{sec:pta} derives their pulsar timing response
and spatial correlation. Section~\ref{sec:discussion} discusses the
observational implications, range of validity, and degeneracy with
Solar System ephemeris errors.

We use natural units with $c=G=1$ and the metric signature
$(-,+,+,+)$. Greek indices $\mu,\nu,\rho,\lambda,\ldots$ denote
four-dimensional spacetime components and run from $0$ to $3$, while
Latin indices $i,j,k,\ldots$ denote three-dimensional spatial
components and run from $1$ to $3$. Parentheses and square brackets
around indices denote symmetrization and anti-symmetrization,
respectively; for example,
$A_{(\mu\nu)}\equiv\frac{1}{2}(A_{\mu\nu}+A_{\nu\mu})$ and
$A_{[\mu\nu]}\equiv\frac{1}{2}(A_{\mu\nu}-A_{\nu\mu})$.
The independent affine connection is denoted by
$\Gamma^{\lambda}{}_{\mu\nu}$, while
$\widehat{\Gamma}^{\lambda}{}_{\mu\nu}$ denotes the Levi--Civita
connection. Hatted geometric quantities are constructed from the
Levi--Civita connection.


\section{Shear Modes in Torsionless Palatini Spacetime}
\label{sec:shear}

GW polarizations are defined by the relative motion of neighboring test particles. In a local detector frame, test particles following Levi--Civita geodesics have a spatial separation vector $\eta^i$ satisfying
\begin{equation}
    \frac{\rmd^2\eta^i}{\rmd t^2}
    =
    -\widehat{R}^{i}{}_{0j0}\eta^j,
    \label{eq:riemann-deviation}
\end{equation}
where $t$ is the coordinate time in the local detector frame and
$\widehat{R}^{i}{}_{0j0}$ is the electric part of the Riemann tensor
constructed from the Levi--Civita connection. For a plane wave propagating along the $+z$ direction, the symmetry of $\widehat{R}_{i0j0}$ gives the conventional six-mode tidal matrix
\begin{equation}
\widehat{R}^{i}{}_{0j0}
=
\begin{pmatrix}
P_4+P_6 & P_5 & P_2\\
P_5 & -P_4+P_6 & P_3\\
P_2 & P_3 & P_1
\end{pmatrix}.
\label{eq:six-mode-matrix}
\end{equation}
Here, $P_1$ denotes the longitudinal mode, $P_2$ and $P_3$ the two vector modes, $P_4$ and $P_5$ the two tensor modes, and $P_6$ the breathing mode. General relativity propagates only the two tensor modes \cite{Eardley:1973zuo}.

Beyond this conventional Riemannian description, the Palatini framework treats the metric and the affine connection as independent variables. For a Palatini theory, the action can be written as
\begin{equation}
    S
    =
    S_g[g_{\mu\nu},\Gamma^{\lambda}{}_{\mu\nu}]
    +
    S_m[g_{\mu\nu},\Gamma^{\lambda}{}_{\mu\nu},\Psi],
    \label{eq:palatini-action}
\end{equation}
where $S$ is the total action, $S_g$ and $S_m$ denote the
gravitational and matter actions, respectively,
$g_{\mu\nu}$ is the spacetime metric, and $\Psi$ collectively denotes
the matter fields. Here, the square brackets in $S_g$ and $S_m$ indicate functional
dependence on the enclosed fields. For the torsionless case considered here, the
independent connection is symmetric in its lower indices,
$\Gamma^{\lambda}{}_{\mu\nu}
=\Gamma^{\lambda}{}_{\nu\mu}$.
The difference between the independent connection
$\Gamma^{\lambda}{}_{\mu\nu}$ and the Levi--Civita connection
$\widehat{\Gamma}^{\lambda}{}_{\mu\nu}$ is described by the distortion
tensor $N^{\lambda}{}_{\mu\nu}$,
\begin{equation}
\begin{split}
    N^{\lambda}{}_{\mu\nu}
    &\equiv
    \Gamma^{\lambda}{}_{\mu\nu}
    -
    \widehat{\Gamma}^{\lambda}{}_{\mu\nu},
    \\
    Q_{\lambda\mu\nu}
    &\equiv
    \nabla_{\lambda}g_{\mu\nu},
    \qquad
    N^{\lambda}{}_{\mu\nu}
    =
    \frac{1}{2}Q^{\lambda}{}_{\mu\nu}
    -
    Q_{(\mu}{}^{\lambda}{}_{\nu)} .
\end{split}
\label{eq:distortion-nonmetricity}
\end{equation}
Here, $Q_{\lambda\mu\nu}$ is the nonmetricity tensor and
$\nabla_\lambda$ denotes the covariant derivative associated with
the independent connection $\Gamma^{\lambda}{}_{\mu\nu}$. In the torsionless case considered here, the distortion contains no independent torsional contribution, and \Eq{eq:distortion-nonmetricity} expresses it entirely in terms of nonmetricity \cite{hehl1995metric,vitagliano2011dynamics,heisenberg2024review}.

Matter can couple to the independent affine connection through an
explicit dependence of the matter action on
$\Gamma^{\lambda}{}_{\mu\nu}$. This coupling is characterized by the
hypermomentum tensor $H^{\mu\nu}{}_{\lambda}$,
\begin{equation}
    H^{\mu\nu}{}_{\lambda}
    \equiv
    -\frac{2}{\sqrt{-g}}
    \frac{\delta S_m}
    {\delta\Gamma^{\lambda}{}_{\mu\nu}}.
    \label{eq:hypermomentum}
\end{equation}
Here $g\equiv\det(g_{\mu\nu})$. For a localized test body, neglecting dipole and higher multipole
moments, the hypermomentum distribution can be written as
\begin{equation}
    H^{\mu\nu}{}_{\lambda}
    =
    \mathcal H^{\mu\nu}{}_{\lambda}
    \frac{\delta^3[\bm{x}-\bm{X}(t)]}
    {\sqrt{-g}\,u^0},
\end{equation}
where $\bm{x}$ denotes a generic spatial position,
$\bm{X}(t)$ is the spatial trajectory of the test body,
$\delta^3$ is the three-dimensional Dirac delta distribution,
and $u^0$ is the temporal component of the four-velocity
$u^\mu=\rmd X^\mu/\rmd\tau$, with $X^\mu(\tau)$ denoting the
worldline and $\tau$ the proper time.
The quantity $\mathcal H^{\mu\nu}{}_{\lambda}$ denotes the integrated
hypermomentum of the test body, and
$\mathcal H_{\lambda}\equiv\mathcal H^{00}{}_{\lambda}$
is the corresponding hypermomentum charge. The resulting equation of
motion is
\begin{equation}
    \mathcal H_{\lambda}
    \left(
    \frac{\widehat{D}u^{\mu}}{\rmd\tau}
    +
    N^{\mu}{}_{\alpha\beta}
    u^{\alpha}u^{\beta}
    +
    N^{\rho}{}_{\alpha\beta}
    u^{\alpha}u^{\beta}u_{\rho}u^{\mu}
    \right)
    =0,
    \label{eq:particle-motion}
\end{equation}
where $\widehat D/\rmd\tau$ denotes the covariant derivative along the
worldline constructed from the Levi--Civita connection. Within the
monopole approximation adopted, a nonvanishing hypermomentum
charge $\mathcal H_{\lambda}\neq0$ implies that the particle dynamics
is governed by the equation in parentheses
\cite{Puetzfeld:2007hr,puetzfeld2014equations,iosifidis2024motion}.

For two neighboring worldlines, the linearized relative-motion equation in a local detector frame now takes the form
\begin{equation}
    \frac{\rmd^2\eta^i}{\rmd t^2}
    =
    -\mathcal A^{i}{}_{j}\eta^j,
    \qquad
    \mathcal A^{i}{}_{j}
    =
    \widehat{R}^{i}{}_{0j0}
    +
    \partial_jN^{i}{}_{00}.
    \label{eq:palatini-deviation}
\end{equation}
The Riemann term is symmetric, whereas the additional term $\partial_jN^{i}{}_{00}$ need not be symmetric in $i$ and $j$.For a plane wave propagating along the $+z$ direction, only the derivative with $j=3$ contributes. The effective tidal matrix can therefore be written as
\begin{equation}
\mathcal A^{i}{}_{j}
=
\begin{pmatrix}
P_4+P_6 & P_5 & P_2+P_7\\
P_5 & -P_4+P_6 & P_3+P_8\\
P_2 & P_3 & P_1
\end{pmatrix}.
\label{eq:eight-mode-matrix}
\end{equation}
The new components $P_7$ and $P_8$ appear in the asymmetric entries,
$\mathcal A^{x}{}_{z}\neq\mathcal A^{z}{}_{x}$ and
$\mathcal A^{y}{}_{z}\neq\mathcal A^{z}{}_{y}$.
They define the shear-$x$ and shear-$y$ modes, respectively \cite{dong2025new}.

The shear modes belong to the vector sector under spatial rotations, but their particle-motion patterns differ from those of the conventional vector modes in Riemannian geometry. For the conventional vector-$x$ mode,
\begin{equation}
    \frac{\rmd^2\eta^x}{\rmd t^2}
    =
    -P_2(t)\eta^z,
    \qquad
    \frac{\rmd^2\eta^z}{\rmd t^2}
    =
    -P_2(t)\eta^x,
    \label{eq:vector-x-motion}
\end{equation}
the transverse and longitudinal motions are coupled symmetrically. For the shear-$x$ mode,
\begin{equation}
    \frac{\rmd^2\eta^x}{\rmd t^2}
    =
    -P_7(t)\eta^z,
    \qquad
    \frac{\rmd^2\eta^y}{\rmd t^2}
    =
    \frac{\rmd^2\eta^z}{\rmd t^2}
    =
    0.
    \label{eq:shear-x-motion}
\end{equation}
An initial longitudinal separation produces transverse relative motion, whereas an initial transverse separation does not excite the reverse longitudinal response. This one-way coupling follows directly from the asymmetry of the effective tidal matrix. The shear-$y$ mode has the similar structure. 

In the gauge-invariant formulation, the shear sector is described by the transverse vectors $\Omega_i$ and $K_i$. For propagating modes with nonzero wavenumber, we define
\begin{equation}
\begin{split}
    \mathcal N^{i}
    &\equiv
    \left.N^{i}{}_{00}\right|_{\mathrm{sh}}
    =
    \Omega^{i}-\partial_0K^{i},
    \qquad
    \partial_i\mathcal N^{i}=0,
    \\
    P_7
    &=
    \partial_3\mathcal N^{1}
    =
    \partial_3\Omega^{1}
    -
    \partial_0\partial_3K^{1},
    \\
    P_8
    &=
    \partial_3\mathcal N^{2}
    =
    \partial_3\Omega^{2}
    -
    \partial_0\partial_3K^{2}.
\end{split}
\label{eq:Ncal-P7P8}
\end{equation}

The notation $\left.\cdots\right|_{\mathrm{sh}}$ means that only the
shear-sector part of a quantity is retained. For the plane wave considered here,
$\partial_0\equiv\partial/\partial t$ and
$\partial_3\equiv\partial/\partial z$.
A spatially homogeneous zero mode does not belong to the propagating shear sector and may be absorbed into the background endpoint motion. \Eq{eq:Ncal-P7P8} shows that the shear amplitudes are spatial derivatives of the transverse field $\mathcal N^i$ along the propagation direction.

The eight components of the effective tidal matrix characterize eight
distinguishable patterns of relative motion. This kinematic
classification does not mean that a generic Palatini theory contains
eight propagating dynamical degrees of freedom. For a specified
Palatini theory, the actual propagating modes, as well as the masses
and dispersion relations of those modes, must be determined from the
corresponding linearized field equations. An explicit Palatini model
supporting the shear modes has been constructed in
Ref.~\cite{dong2025new}.

The transverse field $\mathcal N^i$ introduced in
\Eq{eq:Ncal-P7P8} plays two roles in the present analysis.
Its spatial derivatives along the propagation direction determine the
shear amplitudes $P_7$ and $P_8$, while $\mathcal N^i$ also enters
directly into the equation of motion of matter with a non-negligible
hypermomentum response. The field $\mathcal N^i$ therefore connects
the shear polarizations with the local motions of the pulsar and the
Earth considered in the next section.

\section{Pulsar Timing Response and Correlations of the Palatini Shear Modes}
\label{sec:pta}

In pulsar timing, the pulsar and the Earth act as the emitter and
receiver of the electromagnetic pulse, respectively. Since the photon
frequency measured by an observer at either endpoint depends on the
observer's four-velocity, the shear-induced motions of the two
endpoints can contribute to the pulsar redshift
\cite{rakhmanov2005response,creighton2009pulsar}.

We assume that the electromagnetic field is minimally coupled to the
physical metric \cite{rigouzzo2023coupling}. With $A_\mu$ denoting the electromagnetic four-potential and
$F_{\mu\nu}=2\partial_{[\mu}A_{\nu]}$ the electromagnetic field-strength
tensor, the minimally coupled Maxwell action contains no explicit
dependence on the independent affine connection. This assumption applies only to the electromagnetic
sector; the matter degrees of freedom forming the Earth and pulsar
endpoints may still have a non-negligible effective hypermomentum
response. 

We denote the Earth receiving endpoint by $E$ and place it at
$\bm{x}_E=\bm{0}$. The $a$th pulsar is denoted by $P_a$, with position
\begin{equation}
    \bm{x}_{P_a}=L_a\hat{\bm{n}}_a,
    \qquad
    \hat{\bm{n}}_a\cdot\hat{\bm{n}}_b=\cos\zeta,
    \label{eq:endpoint-geometry-definition}
\end{equation}
where $a$ and $b$ label pulsars, $L_a$ is the distance to pulsar
$a$, $\hat{\bm n}_a$ is the unit vector pointing from the Earth to
pulsar $a$, and $\zeta$ is the angular separation between pulsars $a$ and $b$. 
The electromagnetic pulse propagates from the pulsar to the Earth along the direction $-\hat{\bm{n}}_a$.

In general relativity, free-falling particles follow Levi--Civita
geodesics and acquire no additional motion induced by independent connection. 
The situation considered here is different.
If the Earth and pulsar possess non-negligible hypermomentum responses
at the order of interest, the nonmetricity terms generate an additional
endpoint-velocity component beyond the Levi--Civita geodesic motion,
denoted by $\delta u_A^{i,\mathrm{sh}}$, where $A=E,P_a$ labels the
Earth receiving endpoint and the emitting endpoint of pulsar $a$,
respectively. Here and below, the superscript $\mathrm{sh}$ denotes a
quantity induced by the shear sector.

To obtain the explicit form of $\delta u_A^{i,\mathrm{sh}}$, we expand the endpoint four-velocity around the static background four-velocity
$\bar u^\mu=(1,0,0,0)$, and
retain terms to first order in the spatial velocity perturbation.
Thus, $u^0=1$ and $u^i=\delta u_A^{i,\mathrm{sh}}$ at the order considered. After separating the background Levi--Civita geodesic motion, the three terms in \Eq{eq:particle-motion} reduce to

\begin{subequations}
\label{eq:endpoint-linear-terms}
\begin{align}
    \left.
    \frac{\widehat{D}u^{i}}{\rmd\tau}
    \right|_{\mathrm{sh}}
    &=
    \frac{\rmd\,\delta u_A^{i,\mathrm{sh}}}{\rmd t},
    \label{eq:endpoint-linear-first}
    \\
    \left.
    N^{i}{}_{\alpha\beta}
    u^\alpha u^\beta
    \right|_{\mathrm{sh}}
    &=
    \left.N^{i}{}_{00}\right|_{\mathrm{sh}},
    \label{eq:endpoint-linear-second}
    \\
    \left.
    N^{\rho}{}_{\alpha\beta}
    u^\alpha u^\beta
    u_{\rho}u^{i}
    \right|_{\mathrm{sh}}
    &=0 .
    \label{eq:endpoint-linear-third}
\end{align}
\end{subequations}

Substituting these linearized results back into the equation of motion and using the definition
$\mathcal N^i=\left.N^{i}{}_{00}\right|_{\mathrm{sh}}$
introduced in the previous section, we obtain the endpoint equation of motion induced by nonmetricity,

\begin{equation}
    \frac{\rmd\,\delta u_A^{i,\mathrm{sh}}}{\rmd t}
    =
    -\left.N^{i}{}_{00}\right|_{\mathrm{sh}}
    =
    -\mathcal N^i(t_A,\bm{x}_A),
    \qquad
    \label{eq:endpoint-motion-shear}
\end{equation}
Here $t_A$ and $\bm x_A$ denote the coordinate time and spatial
position of endpoint $A$, respectively. In the general-relativistic limit,
$N^{\lambda}{}_{\mu\nu}=0$,
this additional endpoint motion vanishes, while the usual metric GW response remains unchanged.

Integrating \Eq{eq:endpoint-motion-shear} from an
arbitrary reference time $t_0$ gives
\begin{equation}
    \delta u_A^{i,\mathrm{sh}}
    (t_A,\boldsymbol{x}_A)
    =
    \delta u_A^{i,\mathrm{sh}}
    (t_0,\boldsymbol{x}_A)
    -
    \int_{t_0}^{t_A}
    \mathcal N^i(t',\boldsymbol{x}_A)\,\rmd t'.
    \label{eq:endpoint-velocity-integral}
\end{equation}
The time-independent integration constant is absorbed into the
background endpoint velocity and does not contribute to the
nonzero-frequency response considered below.

The photon frequency measured by an observer at endpoint $A$ is
\begin{equation}
    \nu_A=-p_\mu u_A^\mu,
    \qquad
    \label{eq:observed-frequency}
\end{equation}
where $p^\mu$ is the photon four-momentum and $u_A^\mu$ is the four-velocity of the endpoint. In the linearized background, the unperturbed photon four-momentum is taken to be
\begin{equation}
    p_0^\mu
    =
    \nu_0(1,-\hat{\bm{n}}_a),
    \label{eq:background-photon-momentum}
\end{equation}
where $\nu_0$ is the background photon frequency. The four-velocity correction associated with the shear-induced endpoint motion is written as
\begin{equation}
    \delta u_A^{\mu,\mathrm{sh}}
    =
    \left(0,\delta\bm{u}_A^{\mathrm{sh}}\right).
    \label{eq:endpoint-velocity-perturbation}
\end{equation}
Substituting Eqs.~\eqref{eq:background-photon-momentum} and \eqref{eq:endpoint-velocity-perturbation} into \Eq{eq:observed-frequency}, and retaining the contribution linear in the endpoint velocity, gives
\begin{equation}
    \frac{\delta\nu_A^{\mathrm{sh}}}{\nu_0}
    =
    \hat{\bm{n}}_a\cdot
    \delta\bm{u}_A^{\mathrm{sh}}.
    \label{eq:endpoint-frequency-shift}
\end{equation}

We define the shear redshift as the relative frequency shift between the emitting and receiving endpoints,
\begin{equation}
    z_a^{\mathrm{sh}}(t)
    \equiv
    \frac{
    \delta\nu_{P_a}^{\mathrm{sh}}
    -
    \delta\nu_E^{\mathrm{sh}}
    }{\nu_0},
    \label{eq:shear-redshift-definition}
\end{equation}
\Eq{eq:endpoint-frequency-shift} then yields
\begin{equation}
    z_a^{\mathrm{sh}}(t)
    =
    \hat{\bm{n}}_a\cdot
    \left[
    \delta\bm{u}_{P_a}^{\mathrm{sh}}
    (t-L_a,\bm{x}_{P_a})
    -
    \delta\bm{u}_E^{\mathrm{sh}}
    (t,\bm{0})
    \right],
    \label{eq:shear-redshift-endpoints}
\end{equation}
where $t$ is the reception time at the Earth and $t-L_a$ is the corresponding emission time at the pulsar. 
\Eq{eq:shear-redshift-endpoints} shows that the shear contribution to the redshift is determined by the difference between the pulsar and Earth velocity corrections projected along the pulsar line of sight. It represents the additional contribution to the full PTA redshift generated by the nonmetricity-driven endpoint dynamics.

The shear sector is described by the transverse vector field
$\mathcal N^i$, satisfying $\partial_i\mathcal N^i=0$.
For a plane wave propagating in the direction $\hat{\bm q}$,
we introduce two orthonormal basis vectors
$\bm e_B(\hat{\bm q})$, with $B=X,Y$, spanning the plane
perpendicular to $\hat{\bm q}$:
\begin{equation}
    \hat{\bm q}\cdot\bm e_B(\hat{\bm q})=0,
    \qquad
    \bm e_B(\hat{\bm q})\cdot\bm e_{B'}(\hat{\bm q})
    =\delta_{BB'}.
\end{equation}
For convenience, we work in the frequency domain in the following
analysis. We adopt the two-sided Fourier convention
\begin{equation}
    Y(t)
    =
    \int_{-\infty}^{\infty}\rmd f\,
    \widetilde Y(f)
    \mathrm{e}^{\mathrm{i}2\pi f t},
    \qquad
    \omega=2\pi f ,
    \label{eq:fourier-convention}
\end{equation}
for the quantities considered below.
Here, $f$ is the Fourier frequency, $\omega$ is the corresponding
angular frequency, and a tilde
denotes the Fourier transform of the corresponding time-domain
quantity.

The transverse field $\mathcal N^i$ can then be decomposed into its
two polarization components and expanded over frequency and
propagation direction. For propagating modes with $f\neq0$ and real
wavenumber $k(f)\neq0$, we write
\begin{equation}
\begin{split}
    \mathcal N^i(t,\bm{x})
    ={}&\sum_{B=X,Y}\int_{-\infty}^{\infty}\rmd f
    \int_{S^2}\frac{\rmd^2\hat{\bm q}}{4\pi}\,
    \mathcal N_B(f,\hat{\bm q})e_B^i(\hat{\bm q})\\
    &\times
    \exp\!\left\{\mathrm{i}
    \left[\omega t-k(f)\hat{\bm q}\cdot\bm{x}\right]\right\}.
\end{split}
\label{eq:N-plane-wave}
\end{equation}
Here, $S^2$ denotes the unit sphere, and
$\mathcal N_B(f,\hat{\bm q})$ is the Fourier amplitude of the
$B$th transverse component of $\mathcal N^i$ for propagation in the
direction $\hat{\bm q}$.

The corresponding shear amplitudes follow from the relation derived in
\Eq{eq:Ncal-P7P8}. For propagation along the $+z$ direction,
$P_7=\partial_3\mathcal N^1$ and
$P_8=\partial_3\mathcal N^2$. For a general propagation direction
$\hat{\bm q}$, this relation is generalized to the directional
derivative of the transverse field along $\hat{\bm q}$. Acting on the
plane-wave phase in \Eq{eq:N-plane-wave}, this directional derivative
gives a factor $-\mathrm{i}k(f)$. Therefore,
\begin{equation}
    P_B(f,\hat{\bm q})
    =
    -\mathrm{i}k(f)\mathcal N_B(f,\hat{\bm q}).
    \label{eq:P-to-N}
\end{equation}
For a wave propagating along the $+z$ direction, choosing
$\bm e_X=\hat{\bm x}$ and $\bm e_Y=\hat{\bm y}$ gives
$P_X=P_7$ and $P_Y=P_8$, recovering the shear-$x$ and shear-$y$
components introduced above.

Substituting \Eq{eq:N-plane-wave} into the endpoint equation of motion
and integrating over time, the frequency-domain redshift can be written as
\begin{equation}
    \widetilde z_a^{\mathrm{sh}}(f)
    =
    \sum_{B=X,Y}\int_{S^2}
    \frac{\rmd^2\hat{\bm{q}}}{4\pi}\,
    \mathcal R_{aB}^{\mathrm{sh}}(f,\hat{\bm{q}})
    P_B(f,\hat{\bm{q}}),
    \label{eq:shear-redshift-frequency}
\end{equation}
where the single-pulsar response function is
\begin{equation}
    \mathcal R_{aB}^{\mathrm{sh}}(f,\hat{\bm{q}})
    =
    \frac{\hat{\bm{n}}_a\cdot\bm{e}_B(\hat{\bm{q}})}
    {\omega k(f)}
    \left[
    1-\mathrm{e}^{-\mathrm{i}\Phi_a(f,\hat{\bm{q}})}
    \right],
    \qquad
    \Phi_a
    =
    \omega L_a
    +
    k(f)L_a\hat{\bm{q}}\cdot\hat{\bm{n}}_a.
    \label{eq:shear-response-kernel}
\end{equation}
The factor
$\hat{\bm{n}}_a\cdot\bm{e}_B$
projects the shear direction onto the pulsar line of sight. The factor
$1/k(f)$
follows from \Eq{eq:P-to-N}, while
$1/\omega$
arises from the time integration of the endpoint velocity. The two terms in brackets correspond to the Earth and pulsar contributions, respectively. The phase
$\Phi_a$
therefore describes the relative phase between the pulsar and Earth contributions for the same plane-wave mode.

We assume that the shear background is stationary and isotropic. The two shear components are also assumed to be statistically equivalent and mutually uncorrelated. With this two-sided Fourier convention, their second-order
statistics are taken to be
\begin{equation}
    \av{P_B(f,\hat{\bm{q}})P_{B'}^*(f',\hat{\bm{q}}')}
    =4\pi\,\delta_{BB'}\delta(f-f')
    \delta^{(2)}(\hat{\bm{q}}-\hat{\bm{q}}')S_{\mathrm{sh}}(f).
    \label{eq:shear-statistics}
\end{equation}
Here the angular brackets denote an ensemble average, the asterisk
denotes complex conjugation, and $S_{\mathrm{sh}}(f)$ is the
two-sided power spectral density of each shear component. In \Eq{eq:shear-statistics}, $\delta(f-f')$ states that different frequency components are uncorrelated in a stationary background. The factor $\delta^{(2)}(\hat{\bm{q}}-\hat{\bm{q}}')$ expresses the absence
of correlations between distinct propagation directions, while
isotropy is encoded in the absence of any $\hat{\bm{q}}$ dependence in
$S_{\mathrm{sh}}(f)$. The factor $\delta_{BB'}$ expresses the absence of cross
correlations between the two shear components. Since the two shear components share the same power spectrum $S_{\mathrm{sh}}(f)$, no extra angular structure is introduced by selecting a preferred transverse direction.

Substituting \Eq{eq:shear-redshift-frequency} into the two-point function and using \Eq{eq:shear-statistics}, we obtain
\begin{equation}
    \av{\widetilde z_a^{\mathrm{sh}}(f)
    \widetilde z_b^{\mathrm{sh}*}(f')}
    =\delta(f-f')\frac{S_{\mathrm{sh}}(f)}{\omega^2k^2(f)}
    \mathcal I_{ab}(f),
    \label{eq:redshift-two-point-general}
\end{equation}
where
\begin{equation}
\begin{split}
    \mathcal I_{ab}(f)
    ={}&\int_{S^2}\frac{\rmd^2\hat{\bm{q}}}{4\pi}
    \sum_{B=X,Y}
    (\hat{\bm{n}}_a\cdot\bm{e}_B)
    (\hat{\bm{n}}_b\cdot\bm{e}_B)\\
    &\times
    \left(1-\mathrm{e}^{-\mathrm{i}\Phi_a}\right)
    \left(1-\mathrm{e}^{\mathrm{i}\Phi_b}\right).
\end{split}
\label{eq:shear-overlap-integral}
\end{equation}
At this stage, model dependence enters through
$S_{\mathrm{sh}}(f)$ and the dispersion relation $k(f)$, which also
appears in the finite-distance phases in $\mathcal I_{ab}(f)$.
The full finite-distance kernel therefore depends on the specific form
of $k(f)$. In the short-wavelength limit considered below, however, the
normalized angular correlation becomes independent of the specific
forms of $S_{\mathrm{sh}}(f)$ and $k(f)$.

The transverse basis satisfies the completeness relation
\begin{equation}
\begin{split}
    \sum_{B=X,Y}e_B^ie_B^j
    &=\delta^{ij}-q^iq^j,\\
    \sum_{B=X,Y}(\hat{\bm{n}}_a\cdot\bm{e}_B)
    (\hat{\bm{n}}_b\cdot\bm{e}_B)
    &=\cos\zeta-(\hat{\bm{n}}_a\cdot\hat{\bm{q}})
    (\hat{\bm{n}}_b\cdot\hat{\bm{q}})
    \equiv\Pi_{ab}(\hat{\bm{q}}).
\end{split}
\label{eq:transverse-projector}
\end{equation}
The projector $\delta^{ij}-q^iq^j$ removes the component along the propagation direction and retains the transverse plane containing the two shear components. The resulting $\Pi_{ab}(\hat{\bm{q}})$ is a smooth function of $\hat{\bm{q}}$ and is symmetric under the exchange $a\leftrightarrow b$. The full finite-distance kernel, however, also contains the endpoint phases and therefore satisfies the Hermitian relation
\begin{equation}
    \mathcal I_{ba}(f)
    =
    \mathcal I_{ab}^{*}(f).
    \label{eq:Hermitian-overlap}
\end{equation}
It becomes real and symmetric in the short-wavelength limit considered below.

The endpoint factor can then be expanded into four contributions,
\begin{equation}
\begin{aligned}
    \mathcal I_{ab}&=I_0-I_a-I_b+I_{ab}^{P},\\
    I_0&=\int\frac{\rmd^2\hat{\bm{q}}}{4\pi}\,\Pi_{ab},
    &I_a&=\int\frac{\rmd^2\hat{\bm{q}}}{4\pi}\,\Pi_{ab}\mathrm{e}^{-\mathrm{i}\Phi_a},\\
    I_b&=\int\frac{\rmd^2\hat{\bm{q}}}{4\pi}\,\Pi_{ab}\mathrm{e}^{\mathrm{i}\Phi_b},
    &I_{ab}^{P}&=\int\frac{\rmd^2\hat{\bm{q}}}{4\pi}\,
    \Pi_{ab}\mathrm{e}^{-\mathrm{i}(\Phi_a-\Phi_b)}.
\end{aligned}
\label{eq:four-phase-terms}
\end{equation}
The term $I_0$ is the Earth contribution shared by the two timing links. The terms $I_a$ and $I_b$ are cross terms in which one endpoint is the Earth and the other is a pulsar. The term $I_{ab}^{P}$ connects the two different pulsar endpoints.

The Earth term contains no distance phase and can be evaluated directly,
\begin{equation}
\begin{split}
    I_0
    &=\int_{S^2}\frac{\rmd^2\hat{\bm{q}}}{4\pi}
    \left[\cos\zeta-(\hat{\bm{n}}_a\cdot\hat{\bm{q}})
    (\hat{\bm{n}}_b\cdot\hat{\bm{q}})\right]\\
    &=\cos\zeta
-\frac13
\left(\hat{\bm n}_a\cdot\hat{\bm n}_b\right)
=\frac23\cos\zeta.
\end{split}
\label{eq:I0-cosine}
\end{equation}
In the second step, we use the isotropic average of the propagation
direction. Since no preferred spatial direction remains after the
integration over the unit sphere, the rank-two tensor
$\int \rmd^2\hat{\bm q}\,q_iq_j/(4\pi)$ must be proportional to
$\delta_{ij}$. Taking its trace and using
$\hat{\bm q}\cdot\hat{\bm q}=1$ fixes the proportionality coefficient
to $1/3$, giving
\begin{equation}
    \int_{S^2}\frac{\rmd^2\hat{\bm{q}}}{4\pi}\,q_iq_j
    =\frac13\delta_{ij}.
    \label{eq:isotropic-second-moment}
\end{equation}
Thus, the isotropic average removes the dependence on the individual
propagation direction and leaves only
$\hat{\bm n}_a\cdot\hat{\bm n}_b=\cos\zeta$, yielding
$I_0=(2/3)\cos\zeta$.

The remaining three terms contain pulsar-distance phases. We define
\begin{equation}
    \bm{D}_{ab}=L_a\hat{\bm{n}}_a-L_b\hat{\bm{n}}_b,
    \qquad D_{ab}=|\bm{D}_{ab}|,
    \label{eq:pulsar-separation-vector}
\end{equation}
and write representative phase terms as
\begin{equation}
\begin{split}
    I_a&=\mathrm{e}^{-\mathrm{i}\omega L_a}
    \int\frac{\rmd^2\hat{\bm{q}}}{4\pi}\,
    \Pi_{ab}(\hat{\bm{q}})
    \mathrm{e}^{-\mathrm{i} k(f)\hat{\bm{q}}\cdot(L_a\hat{\bm{n}}_a)},\\
    I_{ab}^{P}&=\mathrm{e}^{-\mathrm{i}\omega(L_a-L_b)}
    \int\frac{\rmd^2\hat{\bm{q}}}{4\pi}\,
    \Pi_{ab}(\hat{\bm{q}})
    \mathrm{e}^{-\mathrm{i} k(f)\hat{\bm{q}}\cdot\bm{D}_{ab}}.
\end{split}
\label{eq:oscillatory-terms}
\end{equation}
These integrals contain a smooth angular function multiplied by a plane-wave phase. To display their asymptotic behavior, for an arbitrary spatial vector $\bm y$, we define
\begin{equation}
    \mathcal J(\bm{y})
    \equiv\int_{S^2}\frac{\rmd^2\hat{\bm{q}}}{4\pi}
    \mathrm{e}^{-\mathrm{i} k(f)\hat{\bm{q}}\cdot\bm{y}}
    =j_0(k(f)|\bm{y}|),
    \qquad
    \int\frac{\rmd^2\hat{\bm{q}}}{4\pi}q_iq_j
    \mathrm{e}^{-\mathrm{i} k(f)\hat{\bm{q}}\cdot\bm{y}}
    =-\frac{1}{k^2(f)}\partial_{y_i}\partial_{y_j}\mathcal J(\bm{y}).
    \label{eq:oscillatory-integral-identity}
\end{equation}
Here $j_\ell$ denotes the spherical Bessel function of the first
kind of order $\ell$. Using \Eq{eq:oscillatory-integral-identity},
the angular integrals containing propagation phases can be reduced to
combinations of spherical Bessel functions and their derivatives.
Since the projector $\Pi_{ab}(\hat{\bm{q}})$ defined in
\Eq{eq:transverse-projector} contains only a zeroth-order
scalar part and a second-order tensor part in the propagation
direction, all relevant integrals can be expressed in terms of
$j_0$ and $j_2$.

Substituting these results into
\Eq{eq:four-phase-terms}, the finite-distance angular kernel can
be written as

\begin{equation}
\begin{aligned}
\mathcal I_{ab}(f)
={}&
\frac{2}{3}\cos\zeta
\\
&-
\frac{2}{3}\cos\zeta\,
\mathrm{e}^{-\mathrm{i}\omega L_a}
\left[
j_0\!\left(k(f)L_a\right)
+
j_2\!\left(k(f)L_a\right)
\right]
\\
&-
\frac{2}{3}\cos\zeta\,
\mathrm{e}^{\mathrm{i}\omega L_b}
\left[
j_0\!\left(k(f)L_b\right)
+
j_2\!\left(k(f)L_b\right)
\right]
\\
&+
\mathrm{e}^{-\mathrm{i}\omega(L_a-L_b)}
\Bigg[
\frac{
2j_0\!\left(k(f)D_{ab}\right)
-
j_2\!\left(k(f)D_{ab}\right)
}{3}
\cos\zeta
\\
&\qquad\qquad
+
j_2\!\left(k(f)D_{ab}\right)
\frac{
\left(L_a-L_b\cos\zeta\right)
\left(L_a\cos\zeta-L_b\right)
}{
D_{ab}^{2}
}
\Bigg],
\qquad a\neq b .
\end{aligned}
\label{eq:shear-overlap-bessel}
\end{equation}
The expression above retains the full finite-distance dependence and
satisfies the Hermitian relation in \Eq{eq:Hermitian-overlap}.

We now consider the short-wavelength (distant-pulsar) limit,
\begin{equation}
    |k(f)|L_a\gg1,
    \qquad
    |k(f)|L_b\gg1,
    \qquad
    |k(f)|D_{ab}\gg1 .
\label{eq:far-distance-conditions}
\end{equation}
For relativistic nanohertz waves, these conditions are typically well
satisfied for PTA pulsars and are standard in correlation analyses
\cite{hellings1983upper,Anholm:2008wy,Chamberlin:2011ev}.
For a general Palatini dispersion relation, they instead
provide explicit conditions on $k(f)$ and the pulsar-pair geometry.

Since the spherical Bessel functions decay at large arguments, the
pulsar-dependent contributions are suppressed in this limit,
\begin{equation}
    I_a\simeq0,
    \qquad
    I_b\simeq0,
    \qquad
    I_{ab}^{P}\simeq0 .
\label{eq:pulsar-term-vanish}
\end{equation}
This suppression results from the angular averaging of the rapidly
varying pulsar-distance phases and does not imply that the
pulsar-endpoint response vanishes for an individual propagation
direction. The common Earth contribution contains no such distance-dependent
phase and is therefore not suppressed by the angular average.

Consequently, in the short-wavelength limit, the angular kernel and the normalized ORF for distinct pulsars are
\begin{equation}
    \mathcal I_{ab}^{\mathrm{far}}(\zeta)
    =
    \frac{2}{3}\cos\zeta,
    \qquad
    \Gamma_{ab}^{\mathrm{sh}}(\zeta)
    \equiv
    \frac{3}{2}\mathcal I_{ab}^{\mathrm{far}}(\zeta)
    =
    \cos\zeta,
    \qquad a\neq b ,
    \label{eq:normalized-shear-correlation}
\end{equation}
where the normalization is chosen such that
$\Gamma_{ab}^{\mathrm{sh}}\to1$ as $\zeta\to0$.

The autocorrelation case requires separate treatment. For $a=b$, one
has $\zeta=0$ and $\bm D_{aa}=0$, so that the pulsar--pulsar phase
difference vanishes. Consequently,
\begin{equation}
\begin{aligned}
    I_{aa}^{P}
    &=
    I_0
    =
    \frac{2}{3},
    \\
    I_a
    &=
    \frac{2}{3}
    \mathrm{e}^{-\mathrm{i}\omega L_a}
    \left[
    j_0\!\left(k(f)L_a\right)
    +
    j_2\!\left(k(f)L_a\right)
    \right],
    \qquad
    I_b=I_a^{*},
    \\
    \mathcal I_{aa}
    &=
    I_0-I_a-I_b+I_{aa}^{P}
    \\
    &=
    \frac{4}{3}
    \left\{
    1-
    \left[
    j_0\!\left(k(f)L_a\right)
    +
    j_2\!\left(k(f)L_a\right)
    \right]
    \cos(\omega L_a)
    \right\}.
\end{aligned}
\label{eq:auto-overlap}
\end{equation}
Although the single-pulsar terms are suppressed for
$|k(f)|L_a\gg1$, the pulsar--pulsar term remains finite because
$\bm D_{aa}=0$. Thus,
\begin{equation}
    \mathcal I_{aa}^{\mathrm{far}}
    =
    \frac{4}{3},
\end{equation}
which differs from the $\zeta\to0$ limit of the distinct-pulsar
cross correlation. In the following, we focus on spatial cross
correlations with $a\neq b$.

Substituting the short-wavelength result into
\Eq{eq:redshift-two-point-general}, the redshift two-point function
for distinct pulsars becomes
\begin{equation}
    \av{\widetilde z_a^{\mathrm{sh}}(f)
    \widetilde z_b^{\mathrm{sh}*}(f')}
    \simeq
    \delta(f-f')\frac{2S_{\mathrm{sh}}(f)}{3\omega^2k^2(f)}
    \cos\zeta,
    \qquad a\neq b.
    \label{eq:final-redshift-cross-spectrum}
\end{equation}
\Eq{eq:final-redshift-cross-spectrum} factorizes the cross spectrum
into the frequency-dependent amplitude
$2S_{\mathrm{sh}}(f)/[3\omega^2k^2(f)]$ and the angular factor
$\cos\zeta$.

The timing residual is defined as the time integral of the redshift,
$R_a^{\mathrm{sh}}(t)=\int^t z_a^{\mathrm{sh}}(t')\,\rmd t'$.
Using the Fourier convention adopted above, this gives
\begin{equation}
    \widetilde R_a^{\mathrm{sh}}(f)
    =
    \frac{\widetilde z_a^{\mathrm{sh}}(f)}
    {2\pi\mathrm{i}f}.
    \label{eq:timing-residual-from-redshift}
\end{equation}
The residual cross spectrum therefore acquires an additional
frequency weight $1/(4\pi^2f^2)$ relative to the redshift cross
spectrum, while its angular correlation remains unchanged and is
given by \Eq{eq:normalized-shear-correlation}.

\section{Dipolar structure, ephemeris degeneracy, and discussion}
\label{sec:discussion}

The spatial correlation derived in
\Eq{eq:normalized-shear-correlation} has a pure dipolar structure,
\begin{equation}
    \Gamma^{\mathrm{sh}}(\zeta)
    =
    \cos\zeta,
    \qquad
    \ell=1,
    \label{eq:pure-dipole}
\end{equation}
and therefore contains only the $\ell=1$ multipole. This simple
angular structure has important consequences for its identification
in PTA data.

In PTA timing analysis, an error $\delta\bm r_{\mathrm{eph}}(t)$ in
the Earth position used for the transformation to the Solar System
barycenter induces a timing residual
\cite{Tiburzi:2015kqa,NANOGrav:2020tig,Champion:2010zz,
Caballero:2018lvc}
\begin{equation}
    R_a^{\mathrm{eph}}(t)
    =
    -\hat{\bm n}_a\cdot\delta\bm r_{\mathrm{eph}}(t).
    \label{eq:ephemeris-residual}
\end{equation}
For an isotropic spatial covariance,
\begin{equation}
    \av{\delta\widetilde r_{\mathrm{eph},i}(f)
    \delta\widetilde r_{\mathrm{eph},j}^*(f')}
    =
    \delta(f-f')S_{\mathrm{eph}}(f)\delta_{ij},
    \label{eq:ephemeris-covariance}
\end{equation}
where $S_{\mathrm{eph}}(f)$ denotes the two-sided power spectral
density of the isotropic ephemeris displacement. The resulting cross correlation is
\begin{equation}
    \av{\widetilde R_a^{\mathrm{eph}}(f)
    \widetilde R_b^{\mathrm{eph}*}(f')}
    =
    \delta(f-f')S_{\mathrm{eph}}(f)\cos\zeta,
\end{equation}
and hence
\begin{equation}
    \Gamma^{\mathrm{eph}}_{ab}(\zeta)
    =
    \Gamma^{\mathrm{sh}}_{ab}(\zeta)
    =
    \cos\zeta,
    \qquad a\neq b .
    \label{eq:exact-degeneracy}
\end{equation}
Such dipolar correlations from Solar System ephemeris errors are
well known in PTA analyses
\cite{Tiburzi:2015kqa,Hobbs:2012apa,Guo:2018rpw}.

Despite their different physical origins, the two signals share the
same geometric structure. More generally, any common vector process
whose projection onto pulsar $a$ is proportional to
$\hat{\bm n}_a$ and whose spatial covariance is isotropic produces a
correlation proportional to
$\hat{\bm n}_a\cdot\hat{\bm n}_b=\cos\zeta$.
The shear response and Solar System ephemeris errors both belong to
this class. Their angular correlations are therefore degenerate, and a
dipolar spatial pattern alone cannot determine the physical origin of
the signal.

If both processes are present and are statistically independent, their isotropic dipolar contributions add in the cross spectrum,
\begin{equation}
 C_{ab}^{(\ell=1)}(f)
=
\left[\mathcal A_{\mathrm{sh}}(f)
+\mathcal A_{\mathrm{eph}}(f)\right]\cos\zeta,
    \label{eq:dipole-amplitude-sum}
\end{equation}
where $C_{ab}^{(\ell=1)}(f)$ denotes the dipolar part of the
timing-residual cross-spectral density, with the common factor
$\delta(f-f')$ suppressed. With the conventions used above, the two
dipolar amplitudes are
\begin{equation}
    \mathcal A_{\mathrm{sh}}(f)
    =
    \frac{2S_{\mathrm{sh}}(f)}
    {3\omega^4k^2(f)},
    \qquad
    \mathcal A_{\mathrm{eph}}(f)
    =
    S_{\mathrm{eph}}(f).
    \label{eq:dipole-amplitudes}
\end{equation}
Statistical independence implies that no cross term between the two
processes contributes.

The angular correlation alone constrains only the sum
$\mathcal A_{\mathrm{sh}}+\mathcal A_{\mathrm{eph}}$, so separating the
two contributions requires additional information. A specific
Palatini theory may provide further predictions for the shear signal,
while Solar System ephemeris uncertainties can be independently
constrained by planetary ranging, spacecraft tracking, and other
observations \cite{NANOGrav:2020tig,Caballero:2018lvc}. Such
model-dependent predictions and independent constraints may help
distinguish the two otherwise spatially degenerate contributions.

Additional spatial information may also arise beyond the assumptions
adopted here. The pure-dipole result relies on a stationary and
isotropic shear background and on the short-wavelength approximation
for distinct pulsars. Retaining finite-distance endpoint phases introduces
pulsar-distance and frequency-dependent structure beyond the
universal $\cos\zeta$ correlation, while an anisotropic shear
background can generate additional angular structure
\cite{Mingarelli:2013dsa,Gair:2015hra}. The former is already encoded
in the finite-distance kernel derived above, whereas the latter
requires a generalized treatment without the isotropic angular
average.

In summary, we have derived the pulsar timing response of the
shear-$x$ and shear-$y$ modes in torsionless Palatini spacetime.
In the short-wavelength limit, the normalized ORF for distinct pulsars
reduces to the pure dipole
$\Gamma_{ab}^{\mathrm{sh}}(\zeta)=\cos\zeta$.
Since the same angular dependence is produced by isotropic Solar
System ephemeris errors, the angular correlation alone cannot determine
the physical origin of a dipolar signal.


\begin{acknowledgments}
 
YMW is supported by the National Natural Science Foundation of China (Grant No.~12505086).
CL is supported by the National Natural Science Foundation of China (Grant No.~12405074).

\end{acknowledgments}


\bibliography{refs}	
	
\end{document}